\pdfoutput=1
\documentclass{PoS}
\usepackage[T1]{fontenc}
\usepackage[utf8]{inputenc}
\usepackage{microtype}
\usepackage{dsfont}
\usepackage{mathtools}
\usepackage{xfrac}
\usepackage{physics}
\usepackage{siunitx}
\usepackage{tikz}
\usepackage{pgf}

\usepackage{caption}
\usepackage{subcaption}
\newcommand{\fm}{\femto\metre}
\newcommand{\e}[1]{\mathrm{e}^{#1}}

\newcommand{\Id}{\mathds{1}}

\makeatletter
\DeclareFontFamily{OMX}{MnSymbolE}{}
\DeclareSymbolFont{MnLargeSymbols}{OMX}{MnSymbolE}{m}{n}
\SetSymbolFont{MnLargeSymbols}{bold}{OMX}{MnSymbolE}{b}{n}
\DeclareFontShape{OMX}{MnSymbolE}{m}{n}{
    <-6>  MnSymbolE5
   <6-7>  MnSymbolE6
   <7-8>  MnSymbolE7
   <8-9>  MnSymbolE8
   <9-10> MnSymbolE9
  <10-12> MnSymbolE10
  <12->   MnSymbolE12
}{}
\DeclareFontShape{OMX}{MnSymbolE}{b}{n}{
    <-6>  MnSymbolE-Bold5
   <6-7>  MnSymbolE-Bold6
   <7-8>  MnSymbolE-Bold7
   <8-9>  MnSymbolE-Bold8PoS
   <9-10> MnSymbolE-Bold9
  <10-12> MnSymbolE-Bold10
  <12->   MnSymbolE-Bold12
}{}

\let\ldanlge\@undefined
\let\rdanlge\@undefined
\DeclareMathDelimiter{\ldanlge}{\mathopen}%
                     {MnLargeSymbols}{'164}{MnLargeSymbols}{'164}
\DeclareMathDelimiter{\rdanlge}{\mathclose}%
                     {MnLargeSymbols}{'171}{MnLargeSymbols}{'171}
\makeatother

\DeclareDocumentCommand\evsub{ s s m g }
{ 
	\IfNoValueTF{#4}
	{
		\IfBooleanTF{#1}
		{\vphantom{#3}\left\ldanlge\smash{#3}\right\rdanlge} 
		{\left\ldanlge{#3}\right\rdanlge} 
	}
	{
		\IfBooleanTF{#1}
		{
			\IfBooleanTF{#2}
			{\left\ldanlge{#4}\middle\vert{#3}\middle\vert{#4}\right\rdanlge} 
			{\vphantom{#3#4}\left\ldanlge\smash{#4}\middle\vert\smash{#3}\middle\vert\smash{#4}\right\rdanlge} 
		}
		{\vphantom{#3}\left\ldanlge{#4}\middle\vert\smash{#3}\middle\vert{#4}\right\rdanlge} 
	}
}

\usetikzlibrary{decorations.markings}
\usetikzlibrary{decorations.pathreplacing}
\usetikzlibrary{patterns}

\newcommand{\propAfull}[3]{
  \begin{tikzpicture}[baseline=-3pt,scale=#3]
    \begin{scope}[thick]
      \draw (0,0.2) arc [start angle=+90, end angle=-90, radius=0.2];
    \end{scope}
    \fill (0,+0.2) circle [radius=1pt] node [left]  {#1};
    \fill (0,-0.2) circle [radius=1pt] node [left]  {#2};
  \end{tikzpicture}
}

\newcommand{\propAfact}[3]{
  \begin{tikzpicture}[baseline=-3pt,scale=#3]
    \begin{scope}[thick]
      \draw[double,red] (0,0.2) arc [start angle=+90, end angle=-90, radius=0.2];
    \end{scope}
    \fill (0,+0.2) circle [radius=1pt] node [left]  {#1};
    \fill (0,-0.2) circle [radius=1pt] node [left]  {#2};
    \draw[dashed] (0.4,-0.3) -- (0.4,0.3);
  \end{tikzpicture}
}

\newcommand{\propAcorr}[3]{
  \begin{tikzpicture}[baseline=-3pt,scale=#3]
    \begin{scope}[thick]
      \draw[double,red] (0,0.2) -- (0.2-0.05,0.2);
      \draw[very thick] (0.2,+0.16) -- (0.2,+0.24);
      \draw             (0.2,0.2) arc [start angle=+90, end angle=-90, radius=0.2];
      \draw[very thick] (0.2,-0.16) -- (0.2,-0.24);
      \draw[double,red] (0.2-0.05,-0.2) -- (0,-0.2);
    \end{scope}
    \fill (0,+0.2) circle [radius=1pt] node [left]  {#1};
    \fill (0,-0.2) circle [radius=1pt] node [left]  {#2};
    \draw[dashed] (0.2,-0.3) -- (0.2,0.3);
  \end{tikzpicture}
}

\newcommand{\propBfull}[3]{
  \begin{tikzpicture}[baseline=-3pt,scale=#3]
    \begin{scope}[thick]
      \draw (0,0) .. controls (0.1,0.2) and (0.5,0.2) .. (0.6,0);
    \end{scope}
    \fill (0  ,0) circle [radius=1pt] node [left]  {#1};
    \fill (0.6,0) circle [radius=1pt] node [right] {#2};
    \draw[dashed] (0.3,-0.3) -- (0.3,0.3);
  \end{tikzpicture}
}

\newcommand{\propBfact}[3]{
  \begin{tikzpicture}[baseline=-3pt,scale=#3]
    \begin{scope}[thick]
      \draw             (0  ,0   ) .. controls (0.06,0.12) and (0.2,0.15) .. (0.3,0.15);
      \draw[double,red] (0.3,0.15) .. controls (0.4,0.15) and (0.54,0.12) .. (0.6,0);
      \fill[white] (0.30,0.12) rectangle (0.34,0.18);
      \draw[very thick] (0.3,0.11) -- (0.3,0.19);
    \end{scope}
    \fill (0  ,0) circle [radius=1pt] node [left]  {#1};
    \fill (0.6,0) circle [radius=1pt] node [right] {#2};
    \draw[dashed] (0.3,-0.3) -- (0.3,0.3);
  \end{tikzpicture}
}

\title{Domain decomposition and multilevel integration for fermions}

\ShortTitle{Domain decomposition and multilevel integration for fermions}

\author{Marco Cè\thanks{Speakers.} \,\thanks{Current affiliation: Helmholtz-Institut Mainz, Johannes Gutenberg-Universität, Mainz, Germany.}\\
       Scuola Normale Superiore, Piazza dei Cavalieri 7, I-56126 Pisa, Italy and INFN, sezione di Pisa, Largo B.\ Pontecorvo 3, I-56127 Pisa, Italy\\
       E-mail: \email{marco.ce@sns.it}}

\author{Leonardo Giusti\\
       Dipartimento di Fisica, Università di Milano--Bicocca, and INFN, sezione di Milano--Bicocca, Piazza della Scienza 3, I-20126 Milano, Italy\\
       E-mail: \email{leonardo.giusti@cern.ch}}

\author{Stefan Schaefer\addtocounter{footnote}{-2}\footnotemark\\
       John von Neumann Institute for Computing (NIC), DESY, Platanenallee 6, D-15738 Zeuthen, Germany\\
       E-mail: \email{stefan.schaefer@desy.de}}

\makeatletter
	\global\@speakertrue
	\global\setbox\@firstaubox
	\hbox{{\let\thanks\@gobble
		\let\footnote\@gobble\small
		\rm Stefan Schaefer \& Marco Cè}}%
\makeatother

\abstract{
The numerical computation of many hadronic correlation functions is exceedingly difficult due to the exponentially decreasing signal-to-noise ratio with the distance between source and sink. Multilevel integration methods, using independent updates of separate regions in space-time, are known to be able to solve such problems but have so far been available only for pure gauge theory.\\
We present first steps into the direction of making such integration schemes amenable to theories with fermions, by factorizing a given observable via an approximated domain decomposition of the quark propagator. This allows for multilevel integration of the (large) factorized contribution to the observable, while its (small) correction can be computed in the standard way.
\vspace*{0.5cm}
\begin{flushright}
DESY 16-246
\end{flushright}
}

\FullConference{34th annual International Symposium on Lattice Field Theory\\
                 24-30 July 2016\\
                 University of Southampton, UK}

\begin{document}

\section{Introduction}

With the current state-of-the-art techniques, the numerical computation of hadronic correlation functions in Monte Carlo simulations of lattice QCD is characterized by a signal-to-noise ratio decreasing exponentially with the time separation of the sources~\cite{Parisi:1983ae,Lepage:1989hd}. This problem afflicts most $n$-point correlation functions with the notable exception of the propagator of the lightest pseudoscalars. Some relevant examples are: meson propagators with disconnected contribution, baryon two-point functions, static-light propagators, etc.

The root of this problem is in the different behaviour with respect to time separation of the gauge field mean value vs.\ the corresponding variance. This can be understood already at the theoretical level analysing the field-theoretical expectation value of fermionic Wick contractions, as done in Refs~\cite{Parisi:1983ae,Lepage:1989hd}.

Hadron propagators that get contributions only from connected quark diagrams decay with distance $d$ on every single gauge configuration with a factor $\e{-M_\pi d/2}$ for every quark line~\cite{Parisi:1983ae}. The simplest example of this is the propagator of a isospin-triplet pseudoscalar meson, to which only the  connected Wick contraction $W_{P_c}$ in Eq.~\eqref{eq:singlet_prop} contributes. Another case is the propagator of the nucleon, which get contributions from two connected Wick contractions\footnote{
  See Ref.~\cite{Ce:2016idq} for the definitions of the fields $N(y)$ and $\bar{N}(x)$ and the Wick contractions $W_{N1}(y,x)$ and $W_{N2}(y,x)$.
}
\begin{equation}
\label{eq:nucleon_prop}
  C_N(y_0,x_0) = \frac{1}{L^3} \sum_{\vec{x},\vec{y}} \ev{ N(y) \bar{N}(x) } = \frac{1}{L^3} \sum_{\vec{x},\vec{y}} \left[ \ev{W_{N1}(y,x)} - \ev{W_{N2}(y,x)} \right]
\end{equation}
With the exception of the lightest pseudoscalar propagator, the gauge field variance decays slower than the expectation value. For instance, in the case of the nucleon propagator in Eq.~\eqref{eq:nucleon_prop} the gauge variance get contributions from diagrams propagating a $3\pi$ state. Therefore, the signal to noise ratio is proportional to
\begin{equation}
  \frac{C_N(y_0,x_0)}{\delta C_N(y_0,x_0)} \propto \exp{ -\left( M_N-\sfrac{3}{2}M_\pi \right) \abs{y-x} } ,
\end{equation}
with $(M_N-\sfrac{3}{2}M_\pi)=\SI{3.7}{\per\fm}$ at the physical point.

The problem is even more severe with propagators that include disconnected contributions, such as the zero-momentum correlator of two isospin-singlet pseudoscalar densities in $N_\text{f}=2$ QCD
\begin{equation}
  C_{P_0}(y_0,x_0) = \frac{1}{L^3} \sum_{\vec{x},\vec{y}} \ev{ P^0(y) P^0(x) } , \qquad P^0(x) = \frac{1}{2} \left( \bar{u}(x)\gamma_5u(x) + \bar{d}(x)\gamma_5d(x) \right) ,
\end{equation}
which is mapped to the two Wick contractions
\begin{equation}
\label{eq:singlet_prop}
  C_{P_0}(y_0,x_0) = \frac{1}{L_3} \sum_{\vec{x},\vec{y}} \left[ \ev{ W_{P_d}(y,x) } - \ev{ W_{P_c}(y,x) } \right] , \qquad
  \begin{aligned}
    W_{P_d}(y,x) &= \tr\{ \gamma_5 D^{-1}(x,x) \} \times \tr \{ \gamma_5 D^{-1}(y,y) \}, \\
    W_{P_c}(y,x) &= \tr\{ D^{-1}(y,x) (D^{-1})^\dagger(y,x) \} .
  \end{aligned}
\end{equation}
The second term is the connected contribution and it is not affected by the signal-to-noise ratio problem. In contrast, the gauge variance of the first term, i.e.\ the disconnected contribution, is not suppressed at all with $d=\abs{y_0-x_0}$. Therefore, the exponentially decaying signal of disconnected Wick contraction has to be extracted in the face of a constant gauge variance.

To make up for this worsening of the signal-to-noise ratio, the number of gauge field configurations need to be increased exponentially to reach the target statistical error at a longer distance. This exponential scaling is difficult to keep up numerically and unsatisfactory from the theoretical point of view.

\section{Multilevel algoritms}

In the cases in which the dynamics is described by a local action, and one is interested in local observables, multilevel Monte Carlo integration algorithms~\cite{Luscher:2001up,Meyer:2002cd,DellaMorte:2008jd,DellaMorte:2010yp} are known to result in an impressive acceleration of the computation. In Section~\ref{sec:two-level}, we describe the simple case of a two-level algorithm for a generic local field theory.

QCD is obviously a local theory, but after integrating out the fermions, the resulting  bosonic theory is 
no longer manifestly local, with 
\begin{itemize}
  \item the contribution of the local quark action  taken into account by the quark determinant;
  \item any product of local quark fields is mapped to Wick contractions, i.e.\ a sum of products of quark propagators.
\end{itemize}
Since both depend on the gauge field of the whole lattice in a non-trivial way, this formulation does not fulfill the 
requirements for a  multilevel Monte Carlo algorithm.

Leaving aside the problem of the quark determinant, that is the subject of another work~\cite{Ce:2016ajy}, in these proceedings we present a method, proposed in Ref.~\cite{Ce:2016idq}, to factorize the gauge-field dependence of the quark propagator. The method is based on approximated factorization, engineered to account for the bulk of the Monte Carlo variance. The exact result is recovered easily with a low-noise global correction term. It follows that observables that depend on the quark propagator can be effectively computed using multilevel algorithms. We tested successfully the cases of the disconnected contribution to the pseudoscalar meson propagator and that of the nucleon propagator.

\subsection{The two-level algorithm}
\label{sec:two-level}

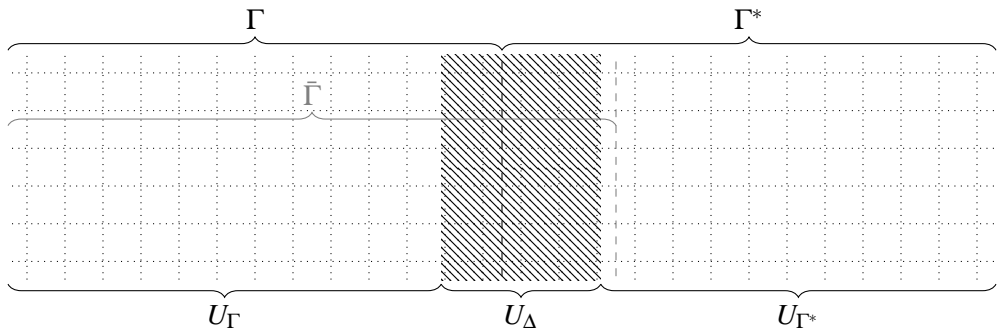
\begin{figure}[b]
  \centering
  \begin{tikzpicture}
    \begin{scope}
      \clip (-6.5,-1.5) rectangle (6.5,+1.5);
      \draw[xshift=0.25cm,yshift=0.25cm,step=0.5,dotted] (-18,-2.5) grid (18,+2.5);
      \draw[dashed]      (0  ,-2.5) -- (0  ,+2.5);
      \draw[dashed,gray] (1.5,-2.5) -- (1.5,+2.5);
      \fill[pattern=north west lines] (-0.8,-2.5) rectangle (1.3,+2.5);
    \end{scope}
    \begin{scope}[yshift=+1.55cm]
      \draw[decorate,decoration={brace,amplitude=5}] (-6.5,0) -- ( 0  ,0) node [midway, above=4] {$\Gamma$};
      \draw[decorate,decoration={brace,amplitude=5}] ( 0  ,0) -- (+6.5,0) node [midway, above=4] {$\Gamma^*$};
    \end{scope}
    \begin{scope}[gray,yshift=+0.55cm]
      \draw[decorate,decoration={brace,amplitude=5}] (-6.5,0) -- (+1.5,0) node [midway, above=4] {$\bar{\Gamma}$};
    \end{scope}
    \begin{scope}[yshift=-1.55cm]
      \draw[decorate,decoration={brace,amplitude=5,mirror}] (-6.5,0) -- (-0.8,0) node [midway, below=4] {$U_{\Gamma}$};
      \draw[decorate,decoration={brace,amplitude=5,mirror}] (-0.8,0) -- (+1.3,0) node [midway, below=4] {$U_{\Delta}$};
      \draw[decorate,decoration={brace,amplitude=5,mirror}] (+1.3,0) -- (+6.5,0) node [midway, below=4] {$U_{\Gamma^*}$};
    \end{scope}
  \end{tikzpicture}
  \caption{Domain decomposition of the lattice in two thick-time-slice regions.}
  \label{fig:dd}
\end{figure}

In this section we describe a simplified domain decomposition of the lattice $\Lambda$ in two regions $\Gamma$ and $\Gamma^*$, such as $\Gamma\cup\Gamma^*=\Lambda$ as pictured in Figure~\ref{fig:dd}. This decomposition is amenable for a two-level integration algorithm. We can decompose the set of gauge links $U$ into the exclusive subsets $U_{\Gamma}$, $U_{\Gamma^*}$ and $U_\Delta$, where $U_i\in\Gamma_i$ and $U_\Delta$ act as a boundary between $U_{\Gamma}$ and $U_{\Gamma^*}$, i.e.\ any path starting in $U_{\Gamma}$ and ending into $U_{\Gamma^*}$ goes through links in $U_\Delta$. We assume that the action can be decomposed as
\begin{equation}
\label{eq:action_decomposition}
  S = S_{\Gamma}[U_{\Gamma},U_\Delta] + S_\partial[U_\Delta] + S_{\Gamma^*}[U_{\Gamma^*},U_\Delta] .
\end{equation}
This decomposition is easily written out for Yang--Mills theory but not for the standard rewriting of QCD in terms of pseudofermions. For an alternative way to write the QCD action in terms of bosonic fields such that the decompositions in Eq.~\eqref{eq:action_decomposition} holds, see Ref.~\cite{Ce:2016ajy}.

For multilevel algorithms to apply, also a factorization of the observable has to be found such that
\begin{equation}
\label{eq:observable_factorization}
  O = O_\text{fact} + O_\text{corr} , \qquad O_\text{fact}[U] = O_{\Gamma}[U_{\Gamma},U_\Delta] \cdot O_{\Gamma^*}[U_{\Gamma^*},U_\Delta] , \quad O_\text{corr}[U] = O[U] - O_\text{fact}[U] .
\end{equation}
This factorization holds exactly, i.e.\ with $O_\text{corr}=0$, for local bosonic observables.
For fermions, we present a method below to find an excellent approximate factorization. 

Then expectation values can be computed by  two-level averaging~\cite{Luscher:2001up,Meyer:2002cd,Ce:2016idq}
\begin{equation}
\label{eq:two-level_average}
  \ev{O} = \ev{ \evsub{O_{\Gamma}}_{\Gamma} \cdot \evsub{O_{\Gamma^*}}_{\Gamma^*} } + \ev{ O_\text{corr} } ,
\end{equation}
where $\evsub{\bullet}_i$ denotes the sublattice expectation value restricted to the gauge field $U_i$
\begin{equation}
  \evsub{O}_i[U_\Delta] = \frac{1}{Z_i} \int \dd{U_i} \e{-S_i[U_i,U_\Delta]} O 
\end{equation}
and the sublattice partition function is fixed requiring $\evsub{\Id}_i=1$.

\subsection{Two-level Monte Carlo}

The field-theoretical two-level expectation value in Eq.~\eqref{eq:two-level_average} has a Monte Carlo realization as follows. Suppose to generate a set of $n_0$ gauge field configurations distributed according to $\e{-S}$. Then, for each one of these \emph{level-$0$} configurations, that we denote with $U^i$ for $i=1,\dotsc,n_0$, we generate a set of $n_1$ configurations in which only the gauge links in $U_{\Gamma}$ and $U_{\Gamma^*}$ are updated. We denote these \emph{level-$1$} configurations with $U^{i,j}$ for $j=1,\dotsc,n_1$. Thanks to the assumption in Eq.~\eqref{eq:action_decomposition} on the form of the action, gauge links belonging to $U_{\Gamma}$ and $U_{\Gamma^*}$ can be updated independently. That is to say, each $U^{i,j}$ represents two different configurations for the observables $O_{\Gamma}$ and $O_{\Gamma^*}$, while it represents a single valid configuration for $O$.

A multilevel-improved estimator for $O$ is given by
\begin{equation}
\label{eq:multilevel_estimator}
  \bar{O} = \frac{1}{n_0} \sum_{i=1}^{n_0} \left\{ \left[ \frac{1}{n_1} \sum_{j=1}^{n_1} O_{\Gamma}[U^{i,j}] \right] \left[ \frac{1}{n_1} \sum_{k=1}^{n_1} O_{\Gamma^*}[U^{i,k}] \right] + \frac{1}{n_1} \sum_{l=1}^{n_1} O_\text{corr}[U^{i,l}] \right\} .
\end{equation}
To argue that the variance of the improved estimator is reduced, assume that the correction $O_\text{corr}$ is suppressed with respect to $O$ by a factor $x$ on every representative gauge field configuration. On one hand, as long as no effects due to the frozen links in $U_{\Delta}$ affect the level-$1$ sampling of $O_\text{corr}$, the gauge variance of the correction is suppressed accordingly to its smallness, such that the statistical error on the estimator of $\ev{O_\text{corr}}$ is $\order{x(n_0\cdot n_1)^{-1/2}}$. On the other hand, $O_{\Gamma}$ and $O_{\Gamma^*}$ fully profit by the two-level averaging, thus the statistical error on the estimator of $\ev{O_\text{fact}}$ is $\order{(n_0\cdot n_1^2)^{-1/2}}$.
In actual cases, significant deviations from this ideal scaling and computational cost factors are likely to be present. Both effects depend on the specific $O$ considered and must be taken into account to estimate the variance reduction.

\section{Quark propagator and locality}

After integrating out analytically the fermions, correlation functions become sum of products of quark propagators that show non-local gauge field dependence. Writing the  Wilson--Dirac operator $D$ in the regions $\Gamma_i$ as a block matrix, its inverse, i.e.\ the quark propagator, is
\begin{equation}
\label{eq:block_inverse}
  D = \begin{pmatrix}
    D_{\Gamma}           & D_{\partial\Gamma} \\
    D_{\partial\Gamma^*} & D_{\Gamma^*}
  \end{pmatrix} \quad \Rightarrow \quad D^{-1} = \begin{pmatrix}
                                            S_{\Gamma}^{-1} & -D_{\Gamma}^{-1} D_{\partial\Gamma} S_{\Gamma^*}^{-1} \\
    -D_{\Gamma^*}^{-1} D_{\partial\Gamma^*} S_{\Gamma}^{-1} &                                     S_{\Gamma^*}^{-1}
  \end{pmatrix} ,
\end{equation}
where \emph{Schur complements}
\begin{equation}
\label{eq:schur_complement}
  S_{\Gamma} = D_{\Gamma} - D_{\partial\Gamma} D_{\Gamma^*}^{-1} D_{\partial\Gamma^*} , \qquad S_{\Gamma^*} = D_{\Gamma^*} - D_{\partial\Gamma^*} D_{\Gamma}^{-1} D_{\partial\Gamma} ,
\end{equation}
are the exact block-inverse of $D$ in regions $\Gamma^*$ and $\Gamma$, respectively. In this block-matrix notation, the diagonal $D_i$ terms depend on the gauge field $U_i$ and eventually on $U_\Delta$. The off-diagonal terms depend on the boundary links $U_\Delta$ only. Clearly, every block of the inverse $D^{-1}$ depends unavoidably on the gauge links of the whole lattice.

Consider the quark propagator $D^{-1}(y,x)$ between a local source in $x$ and sink in $y$. Our goal is to rewrite it as the product of two terms and a correction as in Eq.~\eqref{eq:observable_factorization}. We propose an approximate factorization that is based on the following fact, supported by empirical arguments~\cite{Parisi:1983ae} and numerical evidence: the quark propagator on a given background gauge field configuration is suppressed according to
\begin{equation}
  \norm{ D^{-1}(y,x) } \sim \e{ -\frac{1}{2}M_\pi\abs{y-x} } ,
\end{equation}
where $M_\pi$ is the mass of the lightest pseudoscalar state and $\norm{\bullet}$ is a suitable gauge-invariant norm. Two cases arise: either the source and the sink are in the \emph{same region}, or they are in \emph{different regions}.

\subsection{First case}

\begin{figure}[t]
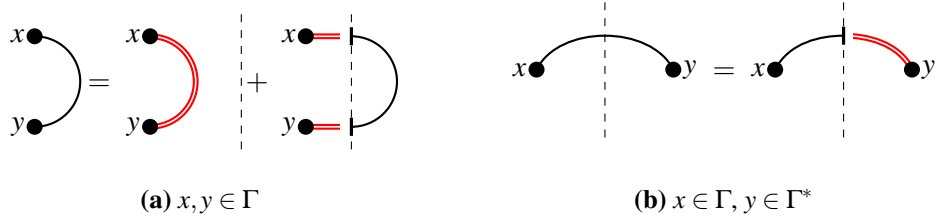

  \centering
  \begin{subfigure}[b]{0.45\textwidth}
  \begin{equation*}
    \propAfull{$x$}{$y$}{3} = \propAfact{$x$}{$y$}{3} + \propAcorr{$x$}{$y$}{3}
  \end{equation*}
  \caption{$x,y\in\Gamma$}\label{fig:quark_prop_I}
  \end{subfigure}
  \begin{subfigure}[b]{0.45\textwidth}
  \begin{equation*}
    \propBfull{$x$}{$y$}{3} = \propBfact{$x$}{$y$}{3} 
  \end{equation*}
  \caption{$x\in\Gamma$, $y\in\Gamma^*$}\label{fig:quark_prop_II}
  \end{subfigure}
  \caption{The factorization of the quark propagator. Black (single) lines are full propagators, red (double) ones are those within a region.}
  \label{fig:quark_prop}
\end{figure}

Suppose that both $x,y\in\Gamma$. The quark propagator is exactly given by
\begin{equation}
\label{eq:quark_prop_I}
  D^{-1}(y,x) = S_{\Gamma}^{-1}(y,x) = D_{\Gamma}^{-1}(y,x) + \left[ D_{\Gamma}^{-1} D_{\partial\Gamma} S_{\Gamma^*}^{-1} D_{\partial\Gamma^*} D_{\Gamma}^{-1} \right](y,x) .
\end{equation}
In this formula, represented graphically in Figure~\ref{fig:quark_prop_I}, the first term depends on the gauge field in the region $\Gamma$ only, while the second term depends on the global gauge field and is the corrections. We argue that the correction is small if $x$ or $y$ are far from the boundary of $\Gamma$. This is supported by the fact that $[D_{\Gamma}^{-1} D_{\partial\Gamma} S_{\Gamma^*}^{-1}](y,\cdot)$ is the exact propagator from a point in $\Gamma^*$ to $y\in\Gamma$, as  can be seen from Eq.~\eqref{eq:block_inverse}. It is thus suppressed on every representative gauge configuration by $\e{-\frac{1}{2}M_\pi d}$, where $d$ is the distance of the external boundary of $\Gamma$ from $y$. An analogous suppression is present for the distance from $x$.

\subsection{Second case}

In the second case, we consider $x\in\Gamma$ and $y\in\Gamma^*$. The factorization in this case is less simple. We can rewrite exactly the propagator as
\begin{equation}
\label{eq:quark_prop_II}
  D^{-1}(y,x) = -\left[ D_{\Gamma^*}^{-1} D_{\partial\Gamma^*} S_{\Gamma}^{-1} \right](y,x) = -\left[ D_{\Gamma^*}^{-1} D_{\partial\Gamma^*} D_{\bar{\Gamma}}^{-1} \right](y,x) + \text{corr} .
\end{equation}
The first equality is pictured in Figure~\ref{fig:quark_prop_II}. Only the $D_{\Gamma}$ factor is local to the gauge field in $\Gamma^*$, while the $S_{\Gamma^*}$ factor depends on the global gauge field. The approximate factorization is obtained by considering a region $\bar{\Gamma}=\Gamma\cup\Delta\Gamma$, where $\Delta\Gamma\subset\Gamma^*$ is a boundary of $\Gamma$ of \emph{thickness} $\Delta$. The approximated quark propagator $-\left[ D_{\Gamma^*}^{-1} D_{\partial\Gamma^*} D_{\bar{\Gamma}}^{-1} \right](y,x)$ is then suitable for a two-level algorithm, provided that the gauge links in $\Gamma^*\cap\bar{\Gamma}=\Delta\Gamma$ are included in the boundary gauge field $U_\Delta$ that is not updated at level-$1$.

It can be shown that the correction in Eq.~\eqref{eq:quark_prop_II} is suppressed on a representative background gauge field according to
\begin{equation}
\label{eq:quark_prop_II_corr}
  \text{corr} = \left[ D_{\Gamma}^{-1} D_{\partial\Gamma} \left( D_{\bar{\Gamma}}^{-1} - S_{\Gamma^*}^{-1} \right) \right](y,x) , \qquad \norm{\text{corr}} \sim \e{-\frac{1}{2}M_\pi\Delta} \cdot \norm{ D^{-1}(y,x) },
\end{equation}
where $\Delta$ is the thickness of the region $\Delta\Gamma$. Eq.~\eqref{eq:quark_prop_II} is the simplest case of an iterative chain of approximations that can be obtained in the general multi-region case, see Ref.~\cite{Ce:2016idq}.

\subsection{Quark line projection}

The quark propagator in Eq.~\eqref{eq:quark_prop_II} is the product of two matrices. When the factorized term is split in $\left[ D_{\Gamma^*}^{-1} D_{\partial\Gamma^*} \right](y,\cdot)$ and $D_{\bar{\Gamma}}^{-1}(\cdot,x)$, the two factors are left with open indices spanning the external boundary of $\Gamma^*$, which in our setup correspond to a tensor with $6L^3$ entries. To reduce the complexity of the two tensors, we ``cut'' the quark line introducing the projection
\begin{equation}
\label{eq:projection}
  P_L = \sum_{i=1}^{N_m} \phi_i \phi_i^\dagger ,
\end{equation}
where $\phi_i$ are $N_m$ orthonormal vectors living on the external boundary of $\Gamma^*$. Then, the quark propagator in Eq.~\eqref{eq:quark_prop_II} is rewritten as
\begin{equation}
\label{eq:quark_prop_II_proj}
  D^{-1}(y,x) = -\sum_{i=1}^{N_m} \left[ D_{\Gamma^*}^{-1} D_{\partial\Gamma^*} \phi_i \right](y) \left[ \phi_i^\dagger D_{\bar{\Gamma}}^{-1} \right](x) + \text{corr} ,
\end{equation}
where the correction term now accounts also for the approximation introduced by the projection. With a suitable choice of projection vectors, the correction is still expected to satisfy Eq.~\eqref{eq:quark_prop_II_corr}.

\section{Domain decomposition and multilevel integration of hadron propagators}

Using the general decomposition of the quark propagator, let us now discuss two applications: the disconnected contribution  of the isospin-singlet propagator $C_{P^0}$ in Eq.~\eqref{eq:singlet_prop} and the nucleon propagator $C_N$ in Eq.~\eqref{eq:nucleon_prop}. In both cases, the application of a two-level algorithm results in an exponential gain in the signal to noise ratio with respect to the standard case. The results for the pion propagator, which does not suffer from a signal-to-noise ratio problem and does not benefit from the multilevel averaging, can be found in Ref.~\cite{Ce:2016idq}.

\subsection{The disconnected contribution}

As has become clear above, the two coordinates in the disconnected
contributions need to reside in the two separate regions $x\in \Gamma$ 
and $y\in \Gamma^*$. According to
the factorization of the quark propagator in Eq.~\eqref{eq:quark_prop_I}, $W_{P_d}$ is  then naturally decomposed as
\begin{equation}
  W_{P_d}(y,x) = W_{P_d}^{(\text{f})}(y,x) + \left[ W_{P_d}^{(\text{r}_1)} + (\Gamma,x)\leftrightarrow(\Gamma^*,y) \right] + W_{P_d}^{(\text{r}_2)}(y,x) ,
\end{equation}
where
\begin{equation}
  W_{P_d}^{(\text{f})}(y,x) = \tr\left\{ \gamma_5 D^{-1}_{\Gamma}(x,x) \right\} \times \tr\left\{ \gamma_5 D^{-1}_{\Gamma^*}(y,y) \right\}
\end{equation}
and the correction $W_{P_d}^{(\text{r})}(y,x)$ is the sum of two terms
\begin{gather}
  W_{P_d}^{(\text{r}_1)} = \tr\left\{ \gamma_5 D^{-1}_{\Gamma}(x,\cdot)\left[D_{\partial\Gamma}D^{-1}D_{\partial\Gamma^*}\right](\cdot,\cdot)D^{-1}_{\Gamma}(\cdot,x) \right\} \times \tr\{ \gamma_5 D^{-1}_{\Gamma^*}(y,y) \} , \label{eq:Wick_disconnected_r1} \\
  \begin{split}
    W_{P_d}^{(\text{r}_2)} =& \tr\left\{ \gamma_5 D^{-1}_{\Gamma}(x,\cdot)\left[D_{\partial\Gamma}D^{-1}D_{\partial\Gamma^*}\right](\cdot,\cdot)D^{-1}_{\Gamma}(\cdot,x) \right\} \\
    \times& \tr\left\{ \gamma_5 D^{-1}_{\Gamma^*}(y,\cdot)\left[D_{\partial\Gamma^*}D^{-1}D_{\partial\Gamma}\right](\cdot,\cdot)D^{-1}_{\Gamma^*}(\cdot,y) \right\} .
  \end{split} \label{eq:Wick_disconnected_r2}
\end{gather}
The contribution term $W_{P_d}^{(\text{f})}(y,x)$ is the product of two factors that depend on the gauge field $U_{\Gamma}\cup U_\Delta$ and $U_{\Gamma^*}\cup U_\Delta$ respectively. This contribution can be estimated with a multilevel algorithm in which the gauge fields $U_{\Gamma}$ and $U_{\Gamma^*}$ are updated independently, with the frozen gauge links region $U_{\Delta}$ as thick as a single time slice.

Since the expressions $W_{P_d}^{(\text{r}_1)}$ and $W_{P_d}^{(\text{r}_2)}$ contain the full quark propagator and depend on the full lattice gauge field, their variance is not reduced by the multilevel algorithm. However, they are proportional to full propagators from either $x$ or $y$ to the boundary between regions. Suppose that $\abs{y-x}=d$ and both $x$ and $y$ are at a distance $d/2$ from the boundary. Then, on every single gauge configuration $W_{P_d}^{(\text{r}_1)}$ is suppressed by a factor $\e{-M_\pi d/2}$ and $W_{P_d}^{(\text{r}_2)}$ by $\e{-M_\pi d}$. The gauge noise of both contributions is expected to be suppressed accordingly, provided the dependence on the frozen gauge links $U_{\Delta}$ is not too strong.

\subsection{The nucleon propagator}
\label{sec:nucleon_fact}

To make the nucleon propagator $C_N(x,y)$  ameanable to the multilevel algorithm, we have to factorize the quark $C_N(x,y)$ using Eq.~\eqref{eq:quark_prop_II_proj} with $x\in\Gamma$ and $y\in\Gamma^*$ including the quark line projectors
\begin{equation}
\label{eq:nucleon_prop_factorized}
  C_N(y_0,x_0) = C_N^{(\text{f})}(y_0,x_0) + C_N^{(\text{r})}(y_0,x_0) = \frac{1}{L^3} \sum_{\vec{x},\vec{y}} \left[ \ev{W_{N1}^{(\text{f})}(y,x)} - \ev{W_{N2}^{(\text{f})}(y,x)} \right] + C_N^{(\text{r})}(y_0,x_0)
\end{equation}
where factorized Wick contractions can be written as
\begin{gather}
  W_{N1}^{(\text{f})}(y,x) = -\sum_{ijk} \mathbb{B}[\xi_j,\xi_i,\xi_k;x]^T \gamma_5 P_- \mathbb{B}[\eta_i,\eta_j,\eta_k;y] , \\
  W_{N2}^{(\text{f})}(y,x) = -\sum_{ijk} \mathbb{B}[\xi_k,\xi_i,\xi_j;x]^T \gamma_5 P_- \mathbb{B}[\eta_i,\eta_j,\eta_k;y] .
\end{gather}
In this expression, the colourless spinor $\mathbb{B}$ denotes the proper contraction of quark fields to create a nucleon
\begin{equation}
  \mathbb{B}[s_i,s_j,s_k;x]_\alpha = \epsilon^{abc} \left\{ [s_i^T]^a(x) C\gamma_5[s_j]^b(x) \right\} [s_k]^c_\alpha(x) ,
\end{equation}
while each quark line is factorized according to Eq.~\eqref{eq:quark_prop_II_proj} in the two contributions
\begin{equation}
  \xi_i^T(x) = \left[ \phi_i^\dagger D^{-1}_{\bar{\Gamma}} \right](x) , \qquad \eta_i(y) = \left[ D^{-1}_{\Gamma^*} D_{\partial\Gamma^*} \phi_i \right](y).
\end{equation}
The first factor in both Wick contractions of Eq.~\eqref{eq:nucleon_prop_factorized} depends only on the gauge field $U_{\Gamma}\cup U_{\Delta}$, while the second factor depends only on the gauge field $U_{\Gamma^*}\cup U_{\Delta}$. This is suitable for the application of the multilevel algorithm. Obviously, the correction term in Eq.~\eqref{eq:nucleon_prop_factorized} depends on the whole lattice gauge field, but it is exponentially suppressed with the thickness of the $\Delta\Gamma$ region and its variance is sufficiently small, so that it can be estimated with a standard Monte Carlo algorithm.

\section{Numerical test for the disconnected pseudoscalar propagator}

We test the two-region domain decomposition in the quenched approximation of
QCD using the Wilson discretization for the gauge action and the Dirac
operator. On the $64\times 24^3$ lattice open boundary conditions in the time direction and periodic
ones in the spatial directions are imposed. The gauge coupling is set with $\beta=6.0$, which corresponds to a
lattice spacing $a\approx\SI{0.093}{\femto\metre}$, and the quark mass with the 
hopping parameter$\kappa=\num{0.1560}$ for the mass-degenerate quark doublet,
which corresponds to a pion mass $M_\pi\approx\SI{455}{\MeV}$.

The disconnected propagagor $W_{P_d}(y,x)$ is estimated stochastically 
\begin{equation}
\label{eq:stochastic_estimator}
  \sum_{\vec{x}} \tr\{ \gamma_5 D^{-1}(x,x) \} \to \frac{1}{n_\text{src}} \sum_{i=1}^{n_\text{src}} \sum_{\vec{x}} \eta_i^\dagger(x) \left[ (\kappa D_\text{hop})^8 D^{-1} \gamma_5 \eta_i \right](x) ,
\end{equation}
where $\eta_i$ are $n_\text{src}=100$ Gaussian random sources defined on the whole spacetime volume and eight powers of $\kappa D_\text{hop}$ to reduce the stochastic estimator noise are employed. $D_\text{hop}$ is defined such that $2\kappa D=\Id-\kappa D_\text{hop}$.

\begin{figure}[p]
  \centering
  \input{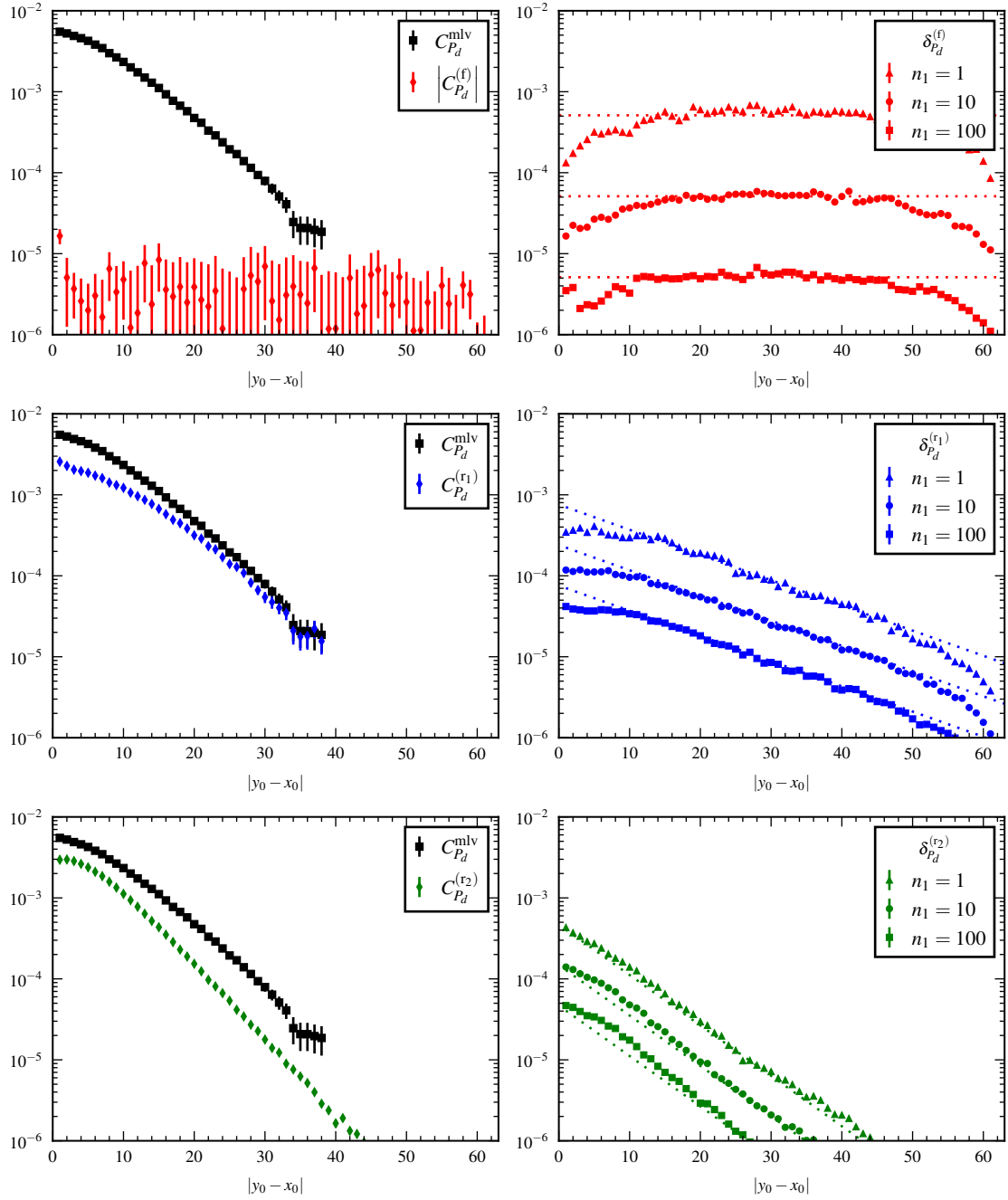}
  \caption{
    Left-column plots: the three contributions on to $C_{P_d}$ are shown, together with the best estimate of the full correlator (the sum of the three), as a function of the time separation $\abs{y_0-x_0}$. Right-column plots: the errors of the various contributions are shown as a function of the time distance for various values of $n_1$.
  }
  \label{fig:singlet}
\end{figure}

To test the factorization of the disconnected pseudoscalar propagator, we decompose the lattice in two domains $\Gamma=\{x\mid x_0\in[0,32]\}$ and $\Gamma^*=\{y\mid y_0\in[33,63]\}$, with  Dirichlet boundary conditions on the block Dirac operators $D_{\Gamma}$ and $D_{\Gamma^*}$, imposed on time slice $x_0=32$ and $x_0=33$ respectively. In this setup, the two contributions to $\sum_{\vec{x},\vec{y}}W_{P_d}^{(\text{f})}(y,x)$,
\begin{equation}
  \sum_{\vec{x}} \tr\{ \gamma_5 D^{-1}_{\Gamma}(x,x) \} \quad\text{and}\quad \sum_{\vec{y}} \tr\{ \gamma_5 D^{-1}_{\Gamma^*}(y,y) \} ,
\end{equation}
have a completely factorized gauge field dependence. They are estimated simultaneously according to Eq.~\eqref{eq:stochastic_estimator} with $n_\text{src}$ solutions of the Dirac equation.

Starting from $n_0=200$ level-$0$ independent gauge field configurations, we generated $n_1=100$ level-$1$ configurations updating independently the gauge links in $U_{\Gamma}$ and $U_{\Gamma^*}$. These gauge fields are defined to be the set of links starting or ending at $x_0<32$ and $x_0>32$ respectively, while spatial links on the $x_0=32$ time slice are assigned to $U_{\Delta}$ and kept fixed during level-$1$ updates. Both $\sum_{\vec{x},\vec{y}}W_{P_d}^{(\text{f})}(y,x)$ and $\sum_{\vec{x},\vec{y}}W_{P_d}(y,x)$ are estimated on a total of $n_0\cdot n_1=\num{20000}$ configurations, and the correction $\sum_{\vec{x},\vec{y}}W_{P_d}^{(\text{r})}(y,x)$ is computed taking the difference. According to the master formula in Eq.~\eqref{eq:multilevel_estimator}, the disconnected contribution $C_{P_d}(y,x)$ is obtained by averaging the factorized Wick contraction independently on level-$1$ configurations, where the correction term is averaged treating the level-$1$ configurations as if they where level-$0$ ones.

The results are shown against the time separation $\abs{y_0-x_0}$ in the six plots of Figure~\ref{fig:singlet}. $x_0\in\Gamma$ and $y_0\in\Gamma^*$ are chosen to be equidistant from the regions boundary $x_0^\text{cut}=32$. Each line shows a different contribution to $C_{P_d}$, with central values on the left and errors on the right. In the first line, the factorized contribution $C_{P_d}^{(\text{f})}$ is shown to be compatible with zero at every distance. Its error is also constant with distance, apart for boundary effects, and clearly decreases $\sim n_1^{-1}$, as a result of the two-level averaging procedure.

In the second line, the first correction $C_{P_d}^{(\text{r1})}$ is obtained as the single-level average of $W_{P_d}^{(\text{r1})}$. Its signal is the dominant contribution to the full correlator for $\abs{y_0-x_0}\gtrsim 15$, while the statistical error decreases as $\e{-M\abs{y_0-x_0}/2}$, as expected by Eq.~\eqref{eq:Wick_disconnected_r1}, but with a ligther effective mass $M=0.14$, which is only $\approx 2/3$ of the pion mass. Moreover, it decreases $\sim n_1^{-1/2}$ as expected from the single-level averaging procedure.

In the third line, the second correction $C_{P_d}^{(\text{r2})}$ is the single-level average of $W_{P_d}^{(\text{r2})}$. Its signal makes a significant contribution to the full correlator for $\abs{y_0-x_0}\lesssim 15$ and the statistical error decreases as $\e{-M\abs{y_0-x_0}}$, which is again as expected from Eq.~\eqref{eq:Wick_disconnected_r2}. Like $C_{P_d}^{(\text{r1})}$, it decreases $\sim n_1^{-1/2}$.

\begin{figure}[t]
  \centering
  \input{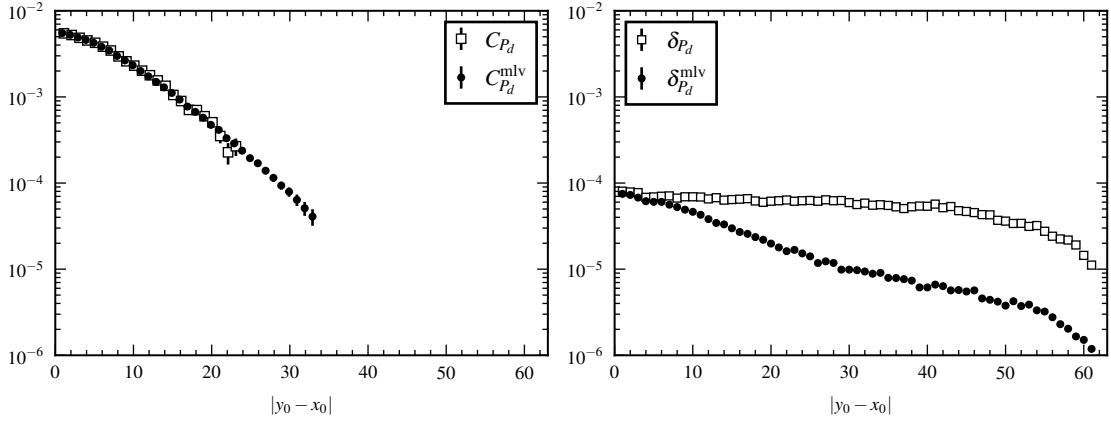}
  \caption{
    The best estimate of $C_{P_d}(y_0,x_0)$ (left) and of its error (right) are shown as a function of the time distance, with and without two-level integration of the factorized contribution. In the latter case the $n_1$ (subset of) configurations, generated for each of the level-0 boundary fields, are treated as if they were correlated level-0 ones. The $n_1$ measures are thus binned together, and the mean and its error are computed as usual by treating the bins as independent.
  }
  \label{fig:singlet_gain}
\end{figure}

Figure~\ref{fig:singlet_gain} shows the best estimate of the full correlation function $C_{P_d}$ obtained from the multilevel algorithm compared to the estimate of the standard case. In the left plot, using the two-level algorithm, the signal is different from zero at the $1\sigma$ level for ten additional time slices. In the right plot is shown how the statical error, which is flat with distance in the standard case, decreases dominated by the first correction contribution up to $\abs{y_0-x_0}=30$. Then, it flattens and it is dominated by the factorized contribution and suppressed by a factor $\sqrt{n_1}=10$ with respect to the standard case. Since the computation of the factorized correlator and its correction require twice the Dirac operator inversion of the standard case, for our choice of $n_1$ the net gain in computational cost is roughly $50$.

\section{Numerical test for the nucleon propagator}

We implemented the factorization of the nucleon propagator described in Section~\ref{sec:nucleon_fact}, as a numerical test of the quark propagator factorization in the second case, i.e Eq.~\eqref{eq:quark_prop_II} or~\eqref{eq:quark_prop_II_proj}. In this case, a overlap region $\Delta\Gamma$ is required for factorization to be a good approximation of the full quark quark propagator. For this reason, we tested the factorization of the nucleon propagator in Eq.~\eqref{eq:nucleon_prop_factorized} decomposing the lattice in the domains $\bar{\Gamma}=\{x\mid x_0\in[0,23+\Delta]\}$ and $\Gamma^*=\{y\mid y_0\in[24,63]\}$, with different choices for the overlap region: $\Delta=8$, $12$ or $16$.

\begin{figure}[p]
  \centering
  \input{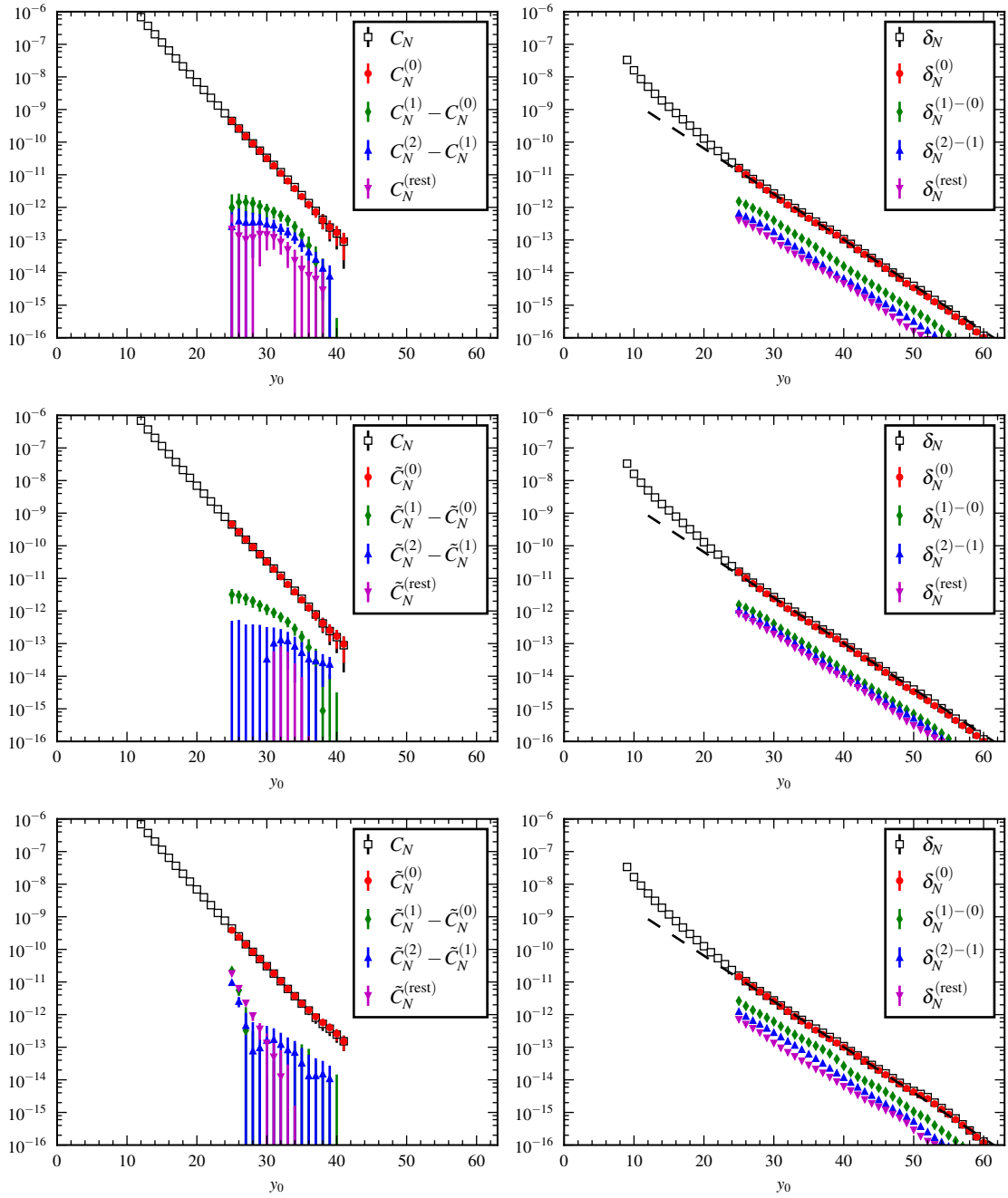}
  \caption{
    Top-line plots: central values (left) and their statistical errors (right) of the five terms appearing on both sides of Eq.~\protect\eqref{eq:nucleon_prop_hierarchical}. For clarity in data are shown only up to $x_0=41$, after which the signal for the correlator is lost. Middle-line plots: analogous results but for a factorized approximation where a projector on the deflation subspace has been inserted to cut the fermion lines. Bottom-line plots: the same but with fermion lines cut by a projector defined via $120$ modes computed by the inverse iteration technique, see main text.
  }
  \label{fig:baryon}
\end{figure}

We expand the nucleon propagator as
\begin{equation}
\label{eq:nucleon_prop_hierarchical}
  C_N = C_N^{(0)} + \left[ C_N^{(1)} - C_N^{(0)} \right] + \left[ C_N^{(2)} - C_N^{(1)} \right] + C_N^{(\text{rest})} ,
\end{equation}
where $C_N^{(i)}$ for $i=0,1,2$ denote the factorized propagators $C_N^{(\text{f})}$ in Eq.~\eqref{eq:nucleon_prop_factorized} with $\Delta=8,12,16$ respectively, while $C_N^{(\text{rest})}=C_N^{(\text{r})}$ for $\Delta=16$.

We first consider the case in which no quark line projection is introduced, i.e.\ the factorization realized by Eq.~\eqref{eq:quark_prop_II}. The plots in the first line of Figure~\ref{fig:baryon} show the results. With the source fixed in $x_0=4$, in the left plot the central value of the propagator is shown against the position of the sink $y_0$. Starting from $x_0=25$, we compute the factorized propagator $C_N^{(0)}$, whose central value is almost indistinguishable from the full propagator. The $[C_N^{(1)}-C_N^{(0)}]$, $[C_N^{(1)}-C_N^{(0)}]$ and $C_N^{(\text{rest})}$ terms are highly suppressed following a clear hierarchy. In the right plot, the statistical error is shown. The error at asymmetrical source-sink separations decays as expected according to $\e{-3M_\pi\abs{y_0-4}/2}$ with $M_\pi=0.215$. This behaviour is observed for the exact propagator as well as the factorized one and all the corrections. The different contributions are suppressed according to the same hierarchy observed for the signal in the left plot.

We then consider the factorization including the projection of quark lines as realized in Eq.~\eqref{eq:quark_prop_II_proj}. We employ two sets of projection vectors. First, the orthonormal vectors that span the deflation subspace as defined in Ref.~\cite{Luscher:2007se}, with local size $N_s=60$. The results are shown in the central line plots of Fig~\ref{fig:baryon}. Second, $N_m=120$ orthonormal vectors constructed by applying $10$ inverse iterations of the Wilson-Dirac operator defined in the domain $\{x\mid x_0\in[24-\Delta,23+\Delta]\}$, with results shown in the bottom line plots of Fig~\ref{fig:baryon}. In both cases, the projection combines well with the factorization: the considerations on the plots in the first line apply, and the results are not affected qualitatively by the introduction of the quark line projection. It is remarkable that the (factorized) baryon propagator is well approximated by the contribution projected on the deflation subspace.

\subsection{Two-level integration}

\begin{figure}[t]
  \centering
  \input{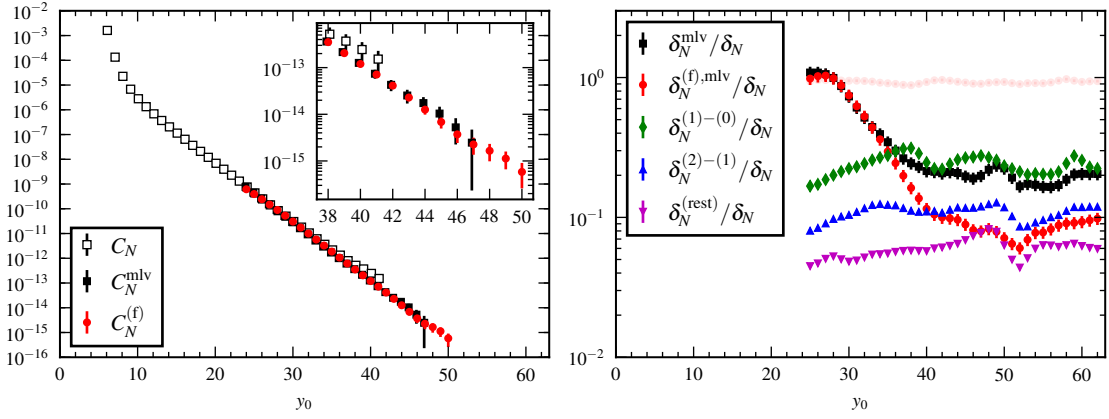}
  \caption{
    Left: best results for $C_N(y_0,x_0)$ with and without two-level integration, and for $C^{(0)}_N(y_0,x_0)$. Right: statistical error of the various contributions of the baryon propagator normalized with the standard average of the exact propagator. For the red factorized contribution, the two-level average is used, with the standard average showed by shadow red points. For completeness we show also in black the statistical error on our best two-level estimate of the exact correlator.
  }
  \label{fig:baryon_ml}
\end{figure}

To test the multilevel integration of the first factorized approximation in Eq.~\eqref{eq:nucleon_prop_hierarchical}, i.e.\ $C_N^{(\text{0})}$ with $\Delta=8$, starting from $n_0=50$ level-$0$ configurations we generate $n_1=20$ level-$1$ gauge fields. The frozen gauge field $U_\Delta$ is composed by the gauge links starting or ending in $\{x\mid x_0\in[16,31]\}$, in order to include the support of the projection vectors in Eq.~\eqref{eq:projection}. The factorized propagator $C_N^{(\text{0})}$ with quark line projected on $N_m=120$ inverse iterations eigenmodes is then two-level averaged on these configurations. The best propagator $C_N^{\text{mlv}}$ is obtained by applying the master formula Eq.~\eqref{eq:multilevel_estimator} to evaluate the factorized approximation. The small correction term can be estimated on the $n_0=1000$ original level-$0$ configurations.

It is worth noting that $\mathbb{B}$ without quark line projection would be a tensor with $\order{(6L^3)^3}$ entries, too complex to be kept in memory. Therefore, the quark line projection, which reduces the number of entries to $\order{N_m^3}$, is fundamental to the computation the two-level average. Only if the number of $n_1$ configurations is small, it may be feasible to compute directly the contractions $[\mathbb{B}^T\gamma_5 P_-\mathbb{B}]$ for each one of the $n_1^2$ combinations of level-$1$ configurations. In Figure~\ref{fig:baryon_ml}, we directly compare $C_N^{\text{mlv}}$ obtained with $n_0\cdot n_1=1000$ total gauge field configurations to the exact propagator $C_N$ computed with the standard algorithm on the same number of configurations. In the left plot is shown how the two-level estimate of the factorized contribution has a signal-to-noise ratio larger than one for $10$ additional time slices with respect to the exact propagator computed with the standard Monte Carlo. When we add the correction to the factorized propagator the gain reduces to $5$ time slices. In the right plot we show in detail the contributions to the statistical error, normalized with respect to the error of the standard Monte Carlo exact propagator. Remarkably, the two-level algorithm results in a gain of one order of magnitude for the statistical error, shown in red, of the factorized propagator at large distances. This gain is the product of multiple factors as analysed in details in Ref.~\cite{Ce:2016idq}. At short distance, the gain is limited by the presence of a region of frozen link in level-$1$ updates. The final error on the best multilevel estimate for the exact propagator is the combination of the errors on the two-level averaged factorized approximation and the most noisy correction shown in green.

\section{Conclusions \& Outlook}

In this talk are presented the results published in Ref.~\cite{Ce:2016idq}, which demonstate that it is possible to factorize the gauge field dependence of the quark propagator in order for multilevel integration scheme to be applicable. The factorization presented here is approximate, but it is designed such that the correction is much smaller and less noisy than the main contribution. In turn, this correction is easily estimated and the exact result recovered with a simple modification of the multilevel algorithm.

The factorized quark propagator is the building block of propagators of phenomenological interest, such as the disconnected meson propagator and the nucleon propagator explicitly considered in this work. We argued analytically and showed numerically in the quenched approximation of QCD that the use of a two-level Monte Carlo algorithm results in an exponential gain in the signal-to-noise ratio. This roughly halves the exponent of the exponential  reduction of the signal-to-noise ratio with the source-sink separation, allowing for significant results at larger distances.

A complete formulation of multilevel integration techniques in QCD beyond the quenched approximation requires an equivalent factorization of quark determinant contribution to the QCD action. This has recently been obtained in Ref.~\cite{Ce:2016ajy}, using techniques related to those presented here.

\bibliographystyle{JHEP}
\bibliography{./biblio.bib}

\end{document}